\pdfoutput=1

\documentclass[11pt]{article}
\usepackage{hyperref} 

\usepackage[final]{acl} 

\usepackage{times}
\usepackage{latexsym}

\usepackage[T1]{fontenc}

\usepackage[utf8]{inputenc}

\usepackage{microtype}

\usepackage{inconsolata}

\usepackage{graphicx}

\usepackage{xcolor}
\usepackage{listings}
\usepackage{array, ragged2e, booktabs} 
\usepackage{multirow}
\usepackage{amsmath, amssymb} 

\definecolor{codegreen}{rgb}{0,0.6,0}
\definecolor{codegray}{rgb}{0.5,0.5,0.5}
\definecolor{codepurple}{rgb}{0.58,0,0.82}
\definecolor{backcolour}{rgb}{0.95,0.95,0.92}

\lstdefinestyle{regex}{
    backgroundcolor=\color{backcolour},
    commentstyle=\color{codegreen},
    keywordstyle=\color{magenta},
    numberstyle=\tiny\color{codegray},
    stringstyle=\color{codepurple},
    basicstyle=\ttfamily\footnotesize,
    breakatwhitespace=false,
    breaklines=true,
    captionpos=b,
    keepspaces=true,
    numbers=left,
    numbersep=5pt,
    showspaces=false,
    showstringspaces=false,
    showtabs=false,
    tabsize=2
}

\title{Automated and Context-Aware Code Documentation Leveraging Advanced LLMs\\}

\author{
  Swapnil Sharma Sarker \\
  Ahsanullah University of Science and Technology \\
  Dhaka, Bangladesh \\
  \texttt{swapnilsharmasarker@gmail.com}
  \And
  Tanzina Taher Ifty \\
  George Mason University \\
  Fairfax, Virginia, USA \\
  \texttt{tifty@gmu.edu}
}

\begin{document}
\maketitle
\begin{abstract}
Code documentation is essential to improve software maintainability and comprehension. The tedious nature of manual code documentation has led to much research on automated documentation generation. Existing automated approaches primarily focused on code summarization, leaving a gap in template-based documentation generation (e.g., Javadoc), particularly with publicly available Large Language Models (LLMs). Furthermore, progress in this area has been hindered by the lack of a Javadoc-specific dataset that incorporates modern language features, provides broad framework/library coverage, and includes necessary contextual information. This study aims to address these gaps by developing a tailored dataset and assessing the capabilities of publicly available LLMs for context-aware, template-based Javadoc generation. In this work, we present a novel, context-aware dataset for Javadoc generation that includes critical structural and semantic information from modern Java codebases. We evaluate five open-source LLMs (including LLaMA-3.1, Gemma-2, Phi-3, Mistral, Qwen-2.5) using zero-shot, few-shot, and fine-tuned setups and provide a comparative analysis of their performance. Our results demonstrate that LLaMA 3.1 performs consistently well and is a reliable candidate for practical, automated Javadoc generation, offering a viable alternative to proprietary systems.
\end{abstract}

\section{Introduction}
\makeatletter
\def\@makefnmark{}  
\renewcommand{\@footnotetext}[1]{%
  \insert\footins{%
    \normalfont\footnotesize
    \interlinepenalty\interfootnotelinepenalty
    \sbox\@tempboxa{#1}%
    \ifdim\wd\@tempboxa<\hsize
      \unhbox\@tempboxa
    \else
      \parbox{\hsize}{#1}%
    \fi
  }%
}
\makeatother
\footnotetext{Code and dataset are available at \url{https://github.com/ineffablekenobi/Documentation-generation-using-LLM}}

Code documentation is a crucial part of software development that bridges the gap between developers, end-users, and future maintainers of a software system. While the fundamental purpose of documentation is to guarantee that the code is comprehensible and accessible, it also acts as a crucial tool for promoting collaboration, boosting productivity, and decreasing technical debt \citep{1}. Without heavily depending on the original code writers, developers can understand the complexities of a code-base, troubleshoot problems, and make well-informed changes with the help of well-structured documentation. A study involving software maintainers highlighted the importance of documentation, revealing that 94.03\% agreed that source code documentation is crucial for object-oriented artifacts \citep{2}.

However, creating and maintaining such documentation is expensive and time-consuming \citep{3}. Many developers fail to document their code consistently or neglect it altogether, leading to technical debt. This is often due to a lack of time, unclear guidelines, or the assumption that the code is self-explanatory \citep{5,6}. Also, developers working in larger teams often lack a standardized approach to documenting code, which causes inconsistency in the style and quality of documentation, which confuses collaborators \citep{7,8}. In order to tackle these issues, developers over the years have often turned to template-based documentation solutions like Javadoc, TSDoc, and OpenAPI specifications, etc., which helped to bring consistency in the style of documentation \citep{5,14,15}. But the inconsistency persists when it comes to the details of the documentation. Some documentation is overly detailed and some is brief, which impacts the readability of the documentation \citep{9,10}. Moreover, large projects make it particularly challenging to manually mark and document code snippets. Furthermore, the documentation often becomes outdated as the development continues maybe because of new features being developed or requirement changes \citep{11,12}. Since one of the main areas where programmers value automation is documentation, an automated solution is therefore quite desirable \citep{4}.

The introduction of Large Language Models (LLM) has revolutionized the field of software development \citep{3,1,18}. Although these models are trained on huge corpus of data from diverse sources, they can be used effectively for tasks like code completion \citep{19}, code generation \citep{20,21}, project planning \citep{22} and documentation generation \citep{3,1,18}. Recently, automated code documentation generation has been in the center of attention in language research, and quite a few advancements have been made in this field. Summarization techniques have already been implemented and evaluated \citep{3} in many different languages with the help of popular datasets like CodeSearchNet \citep{16}, CoDesc \citep{38}. Both text-to-text and LLMs have been used in implementing these solutions. Other studies also focused on comprehensive comparisons between multiple LLMs and evaluated their performance in documentation generation over multiple programming languages \citep{1}. Also, Pandey et al. have explored agent-based approaches (i.e. github copilot) for documentation generation \citep{44}. Moreover, models like GPT-4 \citep{23} have been evaluated in template-based documentation such as Javadoc generation \citep{18}. However, many of these high-performing solutions rely on proprietary, closed-source models or APIs, presenting significant challenges for organizations. Data privacy remains one of the major concerns. Additionally, limitations in customization for specific documentation standards, potential latency issues, restrictive rate limits, ongoing costs, and reliance on external providers hinder their adoption where control and security are crucial. Consequently, while potential is clear, there remains a gap in understanding the capabilities of these models in template-based documentation generation task. To date, no studies have focused on evaluating these open source models for task like Javadoc generation, which require adherence to specific formats and contextual understanding.

Evaluating and fine-tuning publicly available LLMs for Javadoc generation requires a suitable dataset. While large code datasets like CodeSearchNet \citep{16} and CoDesc \citep{38} exist, containing millions of code snippets, they are primarily designed for code summarization, making them ill-suited for generating structured, template-based documentation. Furthermore, these datasets lack contextual information (such as class or package context needed for accurate Javadoc tags) and have limited coverage of modern Java features like lambdas and reactive programming constructs such as Mono and Flux, which are extensively used in Java projects. Finally, there are currently no datasets available for template-based documentation generation tasks like Javadoc. This gap highlights the need for a new dataset specifically for training and evaluating models in Javadoc generation that includes contextual information and modern Java features.

This paper makes the following contributions to address these gaps:
\begin{itemize}
\item Introduction of a new, context-aware dataset for Javadoc generation, covering methods, lambdas, and modern Java features, curated from multiple public codebases.
\item Application of automated and manual filtering techniques to ensure the quality and relevance of the dataset.
\item Systematic evaluation and fine-tuning of five publicly available LLMs (LLaMA-3.1, Gemma-2, Phi-3, Mistral, Qwen-2.5) on the proposed Javadoc generation task.
\item A comparative analysis of model performance across zero-shot, few-shot, and fine-tuned settings, providing insights into their capabilities for automated documentation.
\end{itemize}


\section{Related Works}

Numerous studies have been conducted on code documentation, starting with conventional rule-based techniques and advancing to \textbf{pre-LLM} AI models like LSTM and early Transformer-based methods. Ahmad et al. \citep{26} evaluated the Transformer model, which learns code representation for summarization through a self-attention mechanism. To summarize C\# code snippets, CODE-NN, an LSTM-based model with an attention mechanism, was proposed by Iyer et al. \citep{43}. An early neural attention model for code summarization was presented by Allamanis et al. \citep{45}. It incorporates a dual attention mechanism and convolutional features into a recurrent encoder-decoder architecture. However, they were constrained by low flexibility, poor generalization, limited memory, and an insufficient understanding of the content of the code.

Modern LLMs are increasingly applied to automated code documentation, yet existing studies reveal critical limitations. For instance, Khan et al. \citep{3} used Codex for multi-language documentation, achieving a modest BLEU score of 20.6, while Diggs et al. \citep{27} developed specific prompting strategies and evaluation rubrics for generating comments in legacy systems. Similarly, Geng et al. \citep{30} focused on satisfying developer goals by pre-training models with code-comment pairs. Despite these efforts, performance remains a key issue, with Kneidinger et al. \citep{18} demonstrating that even a powerful proprietary model like \textbf{GPT-4 produces unsatisfactory results} for class-level documentation. Furthermore, a major gap persists: existing research has almost exclusively used proprietary models, overlooking the application of \textbf{open-source LLMs specifically for Javadoc generation}. This leaves developers who require flexible and transparent solutions without a viable alternative to paid, closed-source APIs.

\textbf{Agent-based approaches} have also shown promise in this area. For instance, REPOAGENT is an open-source system that excels at repository-level documentation, though it is limited to Python projects and lacks template support \citep{28}. Similarly, commercial agents like GitHub Copilot have demonstrated significant efficiency gains, saving up to 50\% of the time developers spend on documentation tasks \citep{44}. Although LLM-powered agents have strong capabilities, there are still significant obstacles to overcome before they can be used in the real world.  Data security is still a top priority, particularly when managing private or confidential data without the right protections.  Another challenge is customization, since many LLMs find it difficult to adjust to domain-specific requirements without prompt engineering or fine-tuning.  Furthermore, delay can impact usability, especially in interactive environments where users anticipate prompt responses.  Deploying dependable and safe AI-driven bots requires addressing these constraints.

Several \textbf{existing datasets} are relevant to code documentation generation, but possess limitations for our specific focus on template-based Javadoc. For instance, Hasan et al. introduced \textbf{CoDesc} \citep{38}, a large dataset containing over 4.2 million Java methods paired with natural language descriptions. Despite its size, CoDesc suffers from noise and inconsistencies, lacks necessary contextual information for Javadoc generation, does not provide template-based documentation, and offers limited coverage of modern Java constructs. Similarly, \textbf{CodeXGLUE-CONCODE} \citep{concode} provides Java code snippets and natural language descriptions but shares identical drawbacks when considering template-based documentation needs. \textbf{CodeSearchNet} \citep{16}, introduced by Husain et al., covers multiple programming languages but also exhibits issues like noise, potential duplicate entries, a lack of modern Java features, and a primary focus on function-level summaries rather than structured documentation. While \textbf{The Stack }\citep{thestack} represents a massive collection (over 6TB) of permissively licensed source code across many languages, it is a general code corpus and is not specifically curated or structured for the task of documentation generation.

\section{EXPERIMENT}

This section presents our experimental setup and the corresponding model results.

\subsection{DATASET}

The data collection was guided by several principles to ensure a comprehensive and novel dataset. We prioritized \textbf{diversity} by selecting repositories with varied coding styles and documentation patterns, focusing on projects with \textbf{permissive open-source licenses} (e.g., MIT, Apache 2.0, GPL 3.0) to allow for analysis and redistribution. Projects were selected for their high \textbf{Javadoc prevalence} and significant \textbf{contribution activity}, indicating established documentation practices and wide community adoption. Furthermore, the dataset ensures broad \textbf{framework and library coverage}, including tools like Project Reactor and Spring Boot, and incorporates codebases utilizing \textbf{modern Java features} such as lambdas, generics, and stream APIs to reflect current language usage. Our data was sourced from the following repositories, which were selected to meet the requirements mentioned above.

\begin{table}[ht]
    \centering
    \caption{Public repositories used in the dataset}
    \label{tab:repositories}
    \renewcommand{\arraystretch}{1.2} 
    \footnotesize
    \begin{tabular}{c p{5.5cm}} 
        \toprule
        \textbf{Index} & \textbf{Repository Name} \\
        \midrule
        1  & CitiesAPI \cite{CitiesAPI} \\
        2  & Database-api \cite{DatabaseAPI} \\
        3  & Discord4J \cite{Discord4J} \\
        4  & htmldoclet4jdk8 \cite{htmldoclet4jdk8} \\
        5  & JDA \cite{JDA} \\
        6  & Jestures \cite{Jestures} \\
        7  & Milenage \cite{Milenage} \\
        8  & project-tracking-system-backend-app \cite{ProjectTrackingSystem} \\
        9  & SavageFactions \cite{SavageFactions} \\
        10 & termenu \cite{termenu} \\
        \bottomrule
    \end{tabular}
\end{table}

\begin{figure}[htbp]
\centering
\includegraphics[width=0.88\columnwidth]{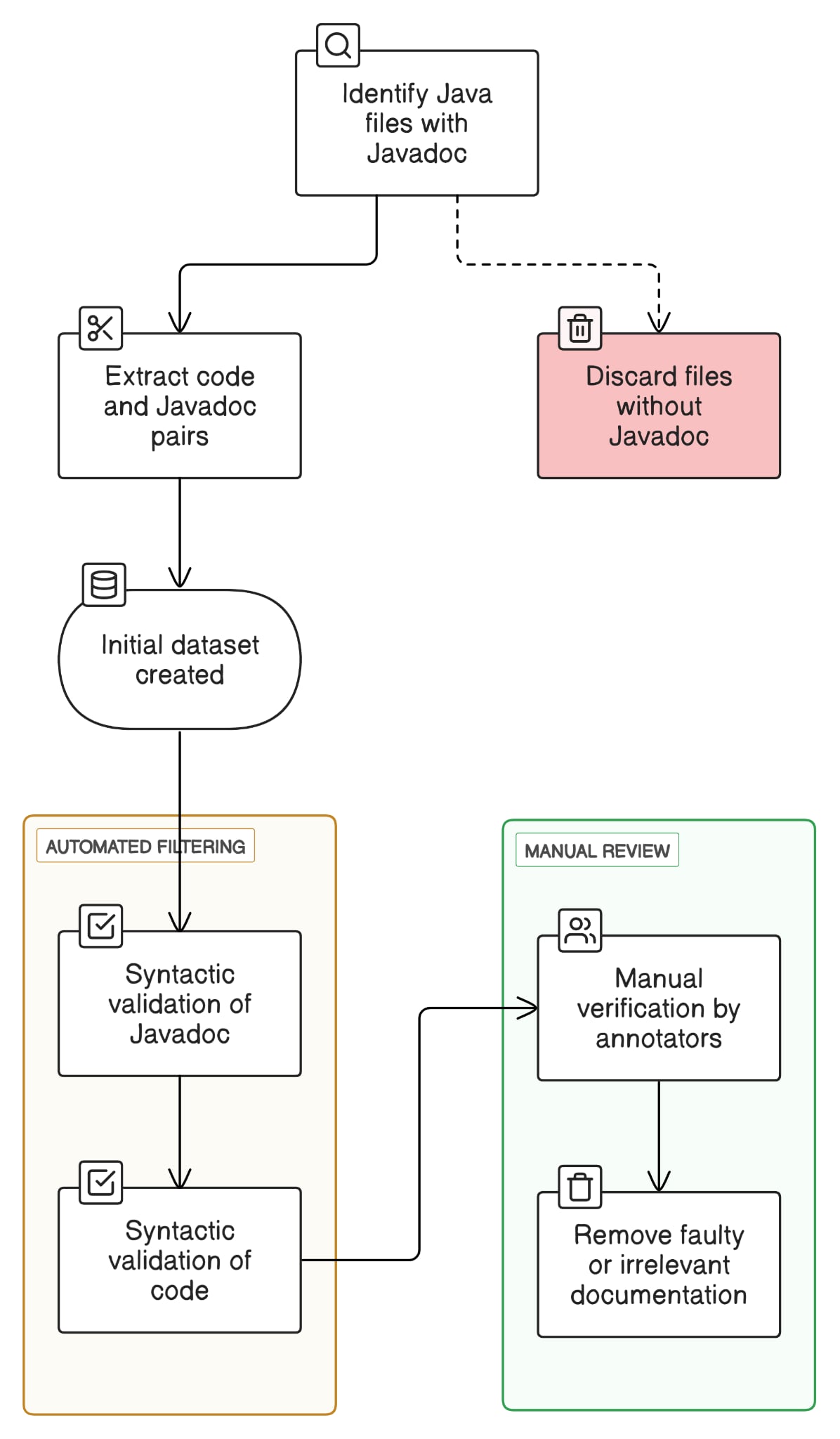} 
\caption{Data Collection and Filtering Pipeline}
\label{fig:data-extract}
\end{figure}

Our data processing pipeline (Figure \ref{fig:data-extract}) begins with scripts identifying .java files containing Javadoc comments (/** ... */) in the selected repositories (Table~\ref{tab:repositories}). Files lacking Javadoc are discarded. From the remaining files, we used a series of regular expressions to parse and extract Javadoc comments, method and class declarations, and package information from the source files. This automated process was followed by syntactic validation to ensure structural integrity. (The specific regular expressions are detailed in Appendix~\ref{sec:appendix_regex}). Essential contextual information, like package and enclosing class names, is captured alongside each code-Javadoc pair, yielding an initial set of roughly 5128 entries.

These pairs then undergo \textbf{automated filtering} using the same patterns to perform syntactic validation. This step verifies the structural integrity of both the Javadoc comments (ensuring proper format) and the code snippets (checking conformance to basic Java syntax). This efficient pattern-based check filters out invalid or incomplete entries, significantly improving dataset quality before manual review.



After automated data filtering, we have conducted a \textbf{thorough review} ensuring correctness and quality of the data. To ensure dataset quality, we instructed annotators to remove entries containing documentation that was faulty, out-of-context, irrelevant or included personal information, and therefore did not accurately describe the corresponding code. Four volunteers, two software engineers, and two academic researchers, generously assisted with the manual verification process. To determine the trustworthiness threshold, we randomly selected 20 samples, distributing 10 of them among four participants who had achieved a 90\% trustworthiness score \cite{41}. The degree of agreement among annotators was assessed using Fleiss’ kappa score \cite{42}, resulting in a value of 0.66, which indicates a substantial level of agreement and helps ensure annotation quality.

 Initially, we have collected 5128 rows of data, which were later filtered based on correctness and relevance. After automated and manual filtering, the dataset contained 3,614 high-quality code-documentation pairs along with their package information. The distribution across training, validation, and test splits is detailed in Table \ref{tab:dataset-distribution}.
\begin{table}[ht]
    \centering
    \caption{Distribution of filtered dataset splits}
    \label{tab:dataset-distribution}
    \renewcommand{\arraystretch}{1.2}
    \begin{tabular}{l r}
        \toprule 
        \textbf{Dataset Type} & \textbf{Number of Samples} \\ \midrule
        Train        & 2,778 \\
        Validation   & 140   \\
        Test         & 696   \\ \midrule
        Total        & 3,614 \\ \bottomrule
    \end{tabular}
\end{table}

Each entry in our final, filtered dataset consists of three key components: the Java code snippet (e.g., a method), its complete Javadoc documentation, and the corresponding package context to aid in understanding dependencies. A concrete example is provided in Appendix~\ref{sec:appendix_dataset_example}.

\subsection{\textbf{Models}}

To achieve optimal outcomes, we fine-tuned five advanced Large Language Models (LLMs) on our dataset: LLaMA-3.1, Gemma-2, Phi-3, Mistral, and Qwen-2.5. Each model employs a decoder-only Transformer architecture with distinct optimizations: \textbf{LLaMA-3.1-8B} utilizes SwiGLU activation and Rotary Positional Embeddings (RoPE) \cite{34}; \textbf{Gemma-2-9B} \cite{35} and \textbf{Phi-3.5-Mini-Instruct} (3.8B) \cite{36} feature RMSNorm, logit soft-capping, and alternating local/global attention; \textbf{Mistral-7B-v0.3} \cite{37} incorporates sliding window attention and grouped-query attention; and \textbf{Qwen-2.5-Coder-3B} \cite{38} includes optimized attention mechanisms and enhanced fine-tuning for long-text generation and instruction following.


\subsection{\textbf{Prompt Engineering}}
Prompts are queries written in a compatible template, so that a model can comprehend what our request is and how it should address the task. Although the specific structure may vary depending on the model, the overall design principles of the prompts are quite similar. Prompts have parts like roles, context, input, etc. \cite{31}. In our study, we designed three distinct types of prompt for different evaluation procedures. Our base prompt format included clear instructions and marked inputs. For \textbf{zero-shot prompting}, we used this base prompt without additional examples. In \textbf{one-shot prompting}, we added a single example to the prompt template, while for \textbf{few-shot prompting}, we carefully handpicked three examples from our dataset to guide the model toward more accurate and desirable responses.

\subsection{\textbf{Evaluation Metrics}}

To assess the quality of the generated documentation, we employed a set of standard, well-established metrics. We used the BLEU score \cite{39} to measure the precision of n-gram overlap between the generated and reference documentation. Additionally, we used several variants of the ROUGE score (R-1, R-2, R-L, and R-Lsum) \cite{40} to evaluate content overlap by assessing recall on unigrams, bigrams, and the longest common subsequence. Detailed definitions and formulas for these metrics can be found in Appendix~\ref{sec:appendix_metrics}.

\subsection{\textbf{Parameter Efficient Training}}
To fine-tune the large language models efficiently under resource constraints, we employed Low-Rank Adaptation (LoRA) \cite{32}, a parameter-efficient training technique. LoRA significantly reduces the number of trainable parameters by freezing the pre-trained weights and injecting smaller, trainable low-rank matrices into the Transformer layers.

We configured LoRA with $\alpha = 16$ to effectively control the influence of low-rank updates on the original weights, balancing responsiveness with parameter stability. Additionally, gradient checkpointing was enabled to reduce memory consumption during back-propagation, allowing for larger batch sizes and efficient GPU memory usage \cite{33}. Table~\ref{tab:trainable-params} illustrates the number of trainable parameters for each model.

\begin{table}[ht]
    \centering
    \caption{Model and number of trainable parameters}
    \label{tab:trainable-params}
    \renewcommand{\arraystretch}{1.2}
    \resizebox{\columnwidth}{!}{%
    \begin{tabular}{p{5cm} r}
        \toprule
        \textbf{Model} & \textbf{Trainable Parameters} \\
        \midrule
        LLaMA-3.1-8B             & 41,943,040 \\
        Mistral-7B-v0.3          & 41,943,040 \\
        Qwen-2.5-Coder-3B        & 29,933,568 \\
        Gemma-2-9B               & 54,018,048 \\
        Phi-3.5-Mini-Instruct    & 29,884,416 \\
        \bottomrule
    \end{tabular}%
    }
\end{table}

We tuned several key hyperparameters, such as the learning rate and weight decay, to ensure stable model convergence. The final training configuration is provided in Appendix~\ref{sec:appendix_hyperparameters}. The use of linear schedulers ensures improved convergence and the weight decay was introduced to prevent possible over-fitting of the models. We always saved the best model (based on validation performance) to ensure that even if overfitting occurs, the selected evaluation model (the best checkpoint) is not affected. All of these models except for Gemma-2-9B were trained on a single P100 GPU in a Kaggle environment. Gemma-2-9B was trained on a single A100 GPU in Google Colab.



\begin{figure*}[ht!] 
\centering
\includegraphics[width=0.95\textwidth]{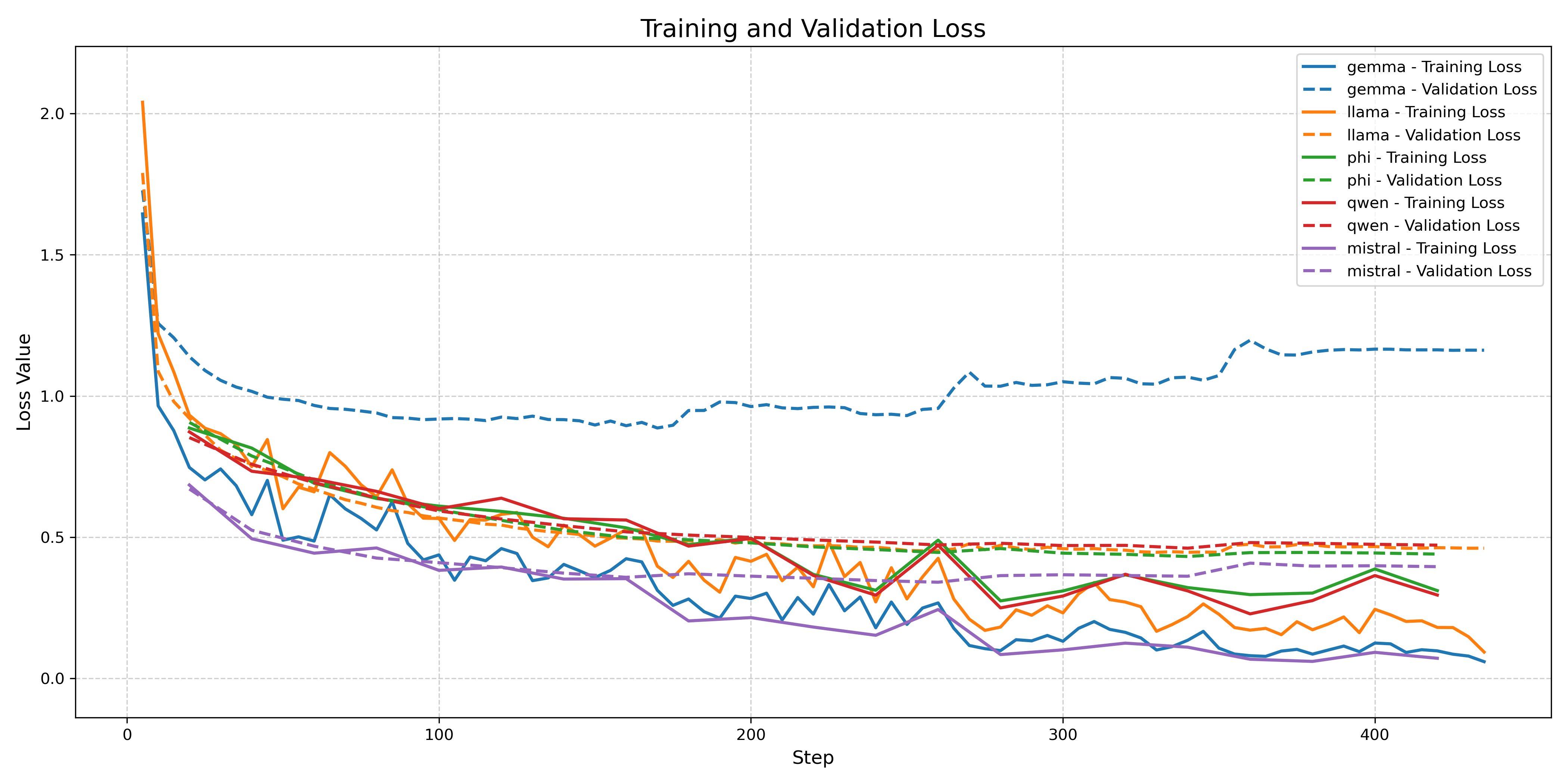} 
\caption{Training and validation loss plot}
\label{fig:training-validation-plot}
\end{figure*}

All models were trained for 5 epochs using a 'steps' evaluation strategy. As shown in Fig.~\ref{fig:training-validation-plot}, the Gemma-2-9B model showed clear signs of overfitting, with its validation loss increasing after 180 steps and remaining significantly higher than other models, possibly due to model complexity or insufficient data. This did not affect the final results, as the best-performing checkpoint was saved. In contrast, LLaMA-3.1-8B's validation loss was the most consistent, stabilizing after 200 steps and suggesting it had reached a point of saturation where further training offered minimal benefit.

Finally, Fig.~\ref{fig:bleu_progression} shows the growth of the BLEU score as a function of the number of steps. It is evident that \textbf{LLaMA-3.1-8B} and \textbf{Mistral-7B-v0.3} performed the best on the validation set, demonstrating consistent performance growth in parallel with the number of steps. Our most efficient model, \textbf{Phi-3.5-Mini-Instruct}, was on par with these models and even outperformed larger models like \textbf{Gemma-2-9B}, though the difference was not substantial.

\subsection{\textbf{Evaluation}}

To assess the performance of the models, we have relied on well-known evaluation metrics, including BLEU and ROUGE scores. The same set of evaluation metrics were consistently implemented to evaluate all the models and then compared to analyze the results and discuss their effectiveness in the documentation generation task.

Firstly, we have focused on the zero-shot evaluation method. In this evaluation technique, we assess the performance of pretrained models without providing them with any examples. The model is provided with simple contexts that tell them their role and the expected output. As shown in Table~\ref{tab:evaluation_results_merged}, \textbf{Qwen-2.5-Coder-3B} showed exceptional performance, outperforming its more complex counterparts in this specific task. The Qwen-2.5 Coder was trained on coding-related tasks like this, making it more suitable for documentation generation. \textbf{LLaMA-3.1-8B} has also shown similar performance despite being a general-purpose model.

\begin{table*}[ht]
    \centering
    \caption{Evaluation results across different settings (Zero-shot, One-shot, Few-shot, and Fine-tuned)}
    \label{tab:evaluation_results_merged}
    \renewcommand{\arraystretch}{1.15}
    \footnotesize
    \begin{tabular}{l l c c c c c}
        \toprule
        \textbf{Setting} & \textbf{Model} & \textbf{BLEU} & \textbf{R1} & \textbf{R2} & \textbf{RL} & \textbf{RLsum} \\
        \midrule
        \multirow{5}{*}{Zero-shot}
        & Gemma-2-9B              & 0.3098 & 0.5522 & 0.2552 & 0.4491 & 0.5429  \\
        & LLaMA-3.1-8B            & 0.3118 & 0.5734 & 0.2667 & 0.4696 & 0.5490  \\
        & Phi-3.5-Mini-Instruct   & 0.2953 & 0.5230 & 0.2261 & 0.4381 & 0.5029  \\
        & \textbf{Qwen-2.5-Coder-3B}       & \textbf{0.3362} & 0.5620 & 0.2770 & 0.4627 & 0.5431  \\
        & Mistral-7B-v0.3         & 0.3118 & 0.5104 & 0.2221 & 0.4342 & 0.5037  \\
        \midrule
        \multirow{5}{*}{One-shot}
        & Gemma-2-9B              & 0.4018 & 0.6270 & 0.2905 & 0.5055 & 0.6140  \\
        & \textbf{LLaMA-3.1-8B}            & \textbf{0.4156} & 0.6428 & 0.2966 & 0.5211 & 0.6222  \\
        & Phi-3.5-Mini-Instruct   & 0.3682 & 0.6016 & 0.2745 & 0.4961 & 0.5830  \\
        & Qwen-2.5-Coder-3B       & 0.4101 & 0.6457 & 0.3243 & 0.5263 & 0.6294  \\
        & Mistral-7B-v0.3         & 0.4135 & 0.6200 & 0.2815 & 0.5147 & 0.6141  \\
        \midrule
        \multirow{5}{*}{Few-shot}
        & Gemma-2-9B              & 0.4422 & 0.6780 & 0.3522 & 0.5524 & 0.6715  \\
        & LLaMA-3.1-8B            & 0.4478 & 0.6677 & 0.3422 & 0.5440 & 0.6579  \\
        & Phi-3.5-Mini-Instruct   & 0.4010 & 0.6279 & 0.2942 & 0.4968 & 0.6217  \\
        & \textbf{Qwen-2.5-Coder-3B}       & \textbf{0.4743} & 0.6852 & 0.3774 & 0.5708 & 0.6783  \\
        & Mistral-7B-v0.3         & 0.4404 & 0.6514 & 0.3294 & 0.5424 & 0.6444  \\
        \midrule
        \multirow{5}{*}{Fine-tuned}
        & Gemma-2-9B              & 0.5318 & 0.8023 & 0.6734 & 0.7782 & 0.7997 \\
        & \textbf{LLaMA-3.1-8B}            & \textbf{0.6606} & 0.9301 & 0.7213 & 0.8125 & 0.8279 \\
        & Phi-3.5-Mini-Instruct   & 0.5987 & 0.8156 & 0.6947 & 0.7986 & 0.8136 \\
        & Qwen-2.5-Coder-3B       & 0.5763 & 0.7936 & 0.6737 & 0.7676 & 0.7908 \\
        & Mistral-7B-v0.3         & 0.6260 & 0.7288 & 0.6372 & 0.7164 & 0.7275 \\
        \bottomrule
    \end{tabular}
\end{table*}

In one-shot and few-shot settings (Table~\ref{tab:evaluation_results_merged}), \textbf{Qwen-2.5-Coder-3B} showed substantial performance gains in the few-shot learning evaluation process by establishing a significant lead over the other models. This illustrates the effect of improved prompts, as the models could use multiple examples as a reference before generating outputs. However, while prompt optimization plays a key role, the training data used to train these models have a significant impact on their understanding of performing a specific task. Therefore, our evaluation results should not be viewed as definitive indicators of a model’s overall capability from a design perspective, but can be considered as reflections of how well a model adapts to this specific task.


After the fine-tuning process, we re-evaluated the models. All models demonstrated substantial performance improvements compared to their pre-fine-tuning results. However, \textbf{LLaMA-3.1-8B} emerged as the top performer, followed by \textbf{Mistral-7B-v0.3}.

\begin{figure}[ht]
    \centering
    \includegraphics[width=\linewidth]{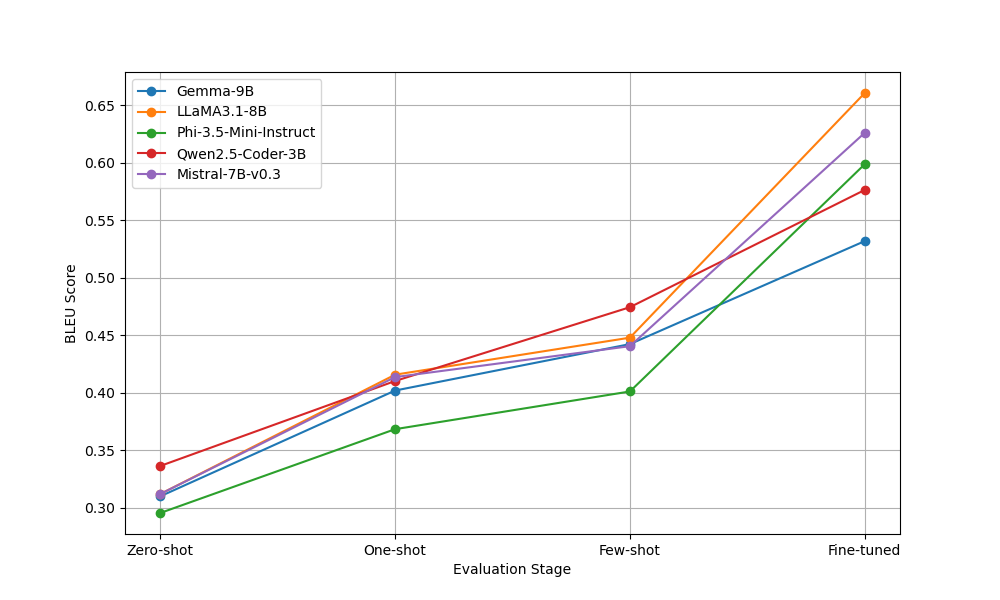} 
    \caption{BLEU score progression over evaluation stages (Zero-shot, One-shot, Few-shot, Fine-tuned)}
    \label{fig:bleu_progression}
\end{figure}

\begin{figure}[ht]
    \centering
    \includegraphics[width=\linewidth]{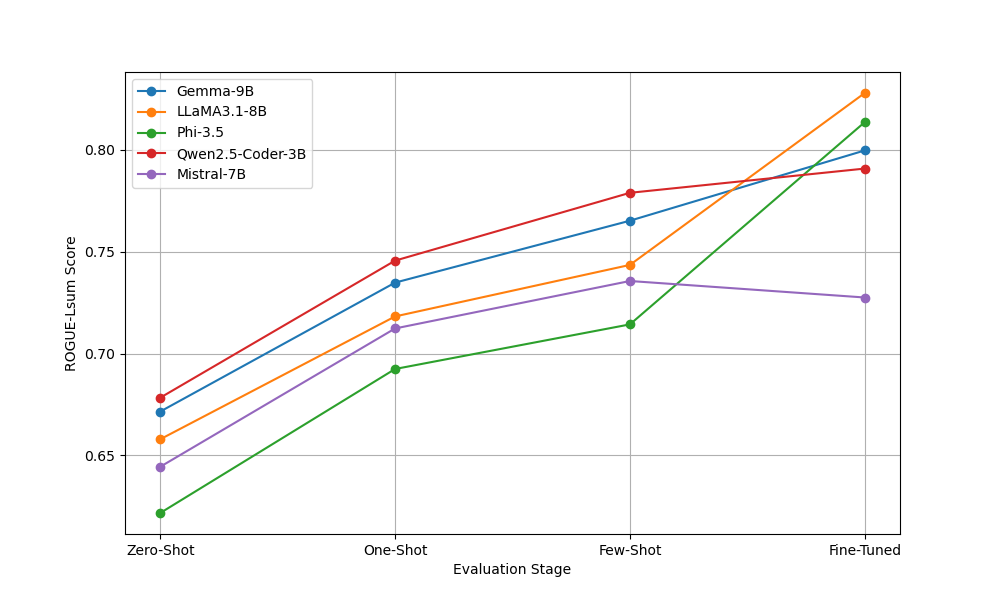} 
    \caption{ROUGE-Lsum score progression over evaluation stages (Zero-shot, One-shot, Few-shot, Fine-tuned)}
    \label{fig:rogue_progression}
\end{figure}

We can see the ROUGE score progression across different evaluation stages illustrated in Fig.~\ref{fig:rogue_progression} (showing ROUGE-Lsum). One interesting observation is \textbf{Mistral-7B-v0.3}’s lower ROUGE scores after fine-tuning compared to some other models, despite its strong BLEU score. This suggests that \textbf{Mistral-7B-v0.3} prioritizes exact replication of patterns learned from fine-tuning data (high BLEU), rather than effectively capturing broader contextual meaning (relatively lower ROUGE). This might imply overfitting to the fine-tuning data structure or a lesser degree of generalization in paraphrasing.

While observing the performance over different stages of evaluation (Fig.~\ref{fig:bleu_progression} and Fig.~\ref{fig:rogue_progression}), it is evident that \textbf{Qwen-2.5-Coder-3B} consistently outperformed every model during the zero-shot, one-shot, and few-shot evaluation stages. However, after fine-tuning, its performance relative to others (especially LLaMA-3.1-8B) was no longer the best. This demonstrates how the pre-training data helped \textbf{Qwen-2.5-Coder-3B} excel initially. After fine-tuning, its performance may have plateaued or the fine-tuning dataset might not have aligned perfectly with its pre-training distributions, causing it to shift away and not maintain its lead.

Meanwhile, \textbf{Phi-3.5-Mini-Instruct}, which previously performed the worst across all earlier evaluations, showed exceptional improvement after fine-tuning. Recall that we mentioned earlier that performance in the initial stages doesn't necessarily represent a model's ability to capture information; this observation serves as evidence of that claim. \textbf{Phi-3.5-Mini-Instruct}’s pre-training data might not have provided enough exposure to the documentation generation task, which could explain its weaker performance in the earlier evaluation stages, but it adapted well during fine-tuning.

Additionally, \textbf{Gemma-2-9B} had the lowest BLEU score after fine-tuning compared to LLaMA and Mistral. However, its ROUGE scores were quite high, particularly R1 and R2, indicating strong content overlap even if exact phrasing (BLEU) differed. This suggests that the model focused on capturing the broader context and semantics rather than simply replicating the reference text structure, which is a positive sign for generating diverse but relevant documentation.

\section{Conclusion}

We introduced a new dataset consisting of reference documentation for methods, lambdas, packages, and class references, designed to provide richer context for fine-tuning publicly available models for Javadoc-style generation. Additionally, we trained models such as LLaMA-3.1-8B, Gemma-2-9B, Phi-3.5-Mini-Instruct, Mistral-7B-v0.3, and Qwen-2.5-Coder-3B on our dataset. Furthermore, we assessed the performance of each model across four different evaluation stages (zero-shot, one-shot, few-shot, and fine-tuned) and measured their effectiveness using BLEU and ROUGE evaluation metrics. Finally, we provided a comprehensive analysis of their performance, highlighting how pre-training influences initial capabilities and how fine-tuning on a targeted documentation generation dataset affects their performance, with LLaMA-3.1-8B showing consistently strong results after fine-tuning. 

\section{Limitations and Risks}
This study has several limitations and risks. \textbf{Resource constraints} limited our dataset size, prevented the fine-tuning of larger model variants, and restricted hyperparameter exploration. The \textbf{dataset diversity} was also insufficient, particularly regarding template-based formats such as TSDoc and JSDoc. Furthermore, our methodology introduced \textbf{fine-tuning risks}, including potential model bias from the training data and a degradation of the model's general-purpose performance. A critical operational risk is the potential for the models to \textbf{generate factually incorrect or misleading documentation (hallucinations)}, which could introduce bugs if trusted by developers without verification.


\section{Acknowledgment}

We express our gratitude to the contributors and maintainers of the open-source models and repositories that have facilitated this research. Specifically, we acknowledge Google for providing \textbf{Gemma-2-9B} under the \href{https://huggingface.co/google/gemma-2-9b/blob/main/LICENSE}{Google AI Model License}, Meta for \textbf{LLaMA-3.1-8B} under the \href{https://huggingface.co/meta-llama/Llama-3.1-8B/blob/main/LICENSE}{Llama 3.1 Community License}, Microsoft for \textbf{Phi-3.5-Mini-Instruct} under the \href{https://huggingface.co/microsoft/Phi-3.5-mini-instruct/blob/main/LICENSE}{MIT License}, Alibaba Qwen Team for \textbf{Qwen2.5-Coder-3B} under the \href{https://huggingface.co/Qwen/Qwen2.5-3B/blob/main/LICENSE}{Qwen Research License}, and Mistral AI for \textbf{Mistral-7B-v0.3} under the \href{https://huggingface.co/mistralai/Mistral-7B-v0.3/blob/main/LICENSE}{Apache 2.0 License}.

Additionally, we thank the volunteers who contributed to the manual verification of the dataset. Finally, we acknowledge the developers of PyTorch, Unsloth, and other open-source libraries used in this study, including Evaluate, Rouge Score, TensorBoard, and gdown, for enabling efficient experimentation and evaluation.

\section{Data availability} 

To facilitate reproducibility and further research, the curated dataset and the code used for model fine-tuning and evaluation are made publicly available. During the anonymous review period, they can be accessed at the following repository: \href{https://anonymous.4open.science/r/automated-documentation-generation-using-llm}{https://anonymous.4open.science/r/automated-documentation-generation-using-llm}. Upon acceptance, the material will be made available via a persistent public repository under a MIT License, CC-BY 4.0 License. As this work utilizes publicly available codebases as its source, our curated and filtered dataset is provided in accordance with open data sharing requirements.

\section{Ethics Statement}

This research follows the principles of the ACM Code of Ethics. Our main goal is to support the software development community by creating and testing open-source tools for automated documentation. We aim to help developers work more efficiently and make software easier to maintain (ACM Code 1.1).

We understand that our work could have negative effects, and we have taken steps to reduce these risks (ACM Code 1.2, 2.5). The biggest risk is that our models could generate documentation that is wrong or confusing (a phenomenon known as "hallucination"). If developers trust this documentation without checking it, it could lead to software bugs, security issues, or a poor understanding of the code. For this reason, we believe these models should be used to assist developers, not to replace them. It is essential that a human developer always reviews the final output.

To be fair and protect privacy (ACM Code 1.4, 1.7), we built our dataset using only public repositories with permissive open-source licenses. We carefully checked the data by hand to find and remove any personal or sensitive information. We also know that the original training data may contain biases, which our models could learn and repeat. While our filtering helps, we acknowledge that the risk of spreading these biases is a limitation of our work.

By making our dataset and code publicly available, we aim to be honest and trustworthy (ACM Code 1.3). This allows the community to be transparent and enables others to reproduce, critique, and build upon our research.

\bibliography{acl_latex}

\clearpage

\appendix
\newpage
\section*{Appendix}

\section{Dataset Example}
\label{sec:appendix_dataset_example}

Below is an example entry from our curated dataset, illustrating the structure which includes the source code snippet, its corresponding Javadoc documentation, and the package context.

\begin{table}[h!]
\centering
\renewcommand{\arraystretch}{1.3}
\begin{tabular}{>{\RaggedRight}p{7cm}}
\toprule
\textbf{Component} \\
\midrule
\textbf{Package:} \\
\texttt{discord4j.core.object} \\
\midrule
\textbf{Code:} \\
\begin{lstlisting}[language=Java, basicstyle=\ttfamily\small]
public Optional<Snowflake> getBotId() {
    return data.botId().toOptional()
               .map(Snowflake::of);
}
\end{lstlisting} \\
\midrule
\textbf{Documentation:} \\
\begin{lstlisting}[language=Java, basicstyle=\ttfamily\small]
/**
 * Gets the id of the bot this role 
 * belongs to, if present.
 *
 * @return The id of the bot this role 
 * belongs to, if present.
 */
\end{lstlisting} \\
\bottomrule
\end{tabular}
\caption{A sample data entry from the curated dataset.}
\label{tab:dataset_example}
\end{table}

\section{Regular Expressions for Data Extraction}
\label{sec:appendix_regex}
\begin{figure}[h!]
    \centering
    \includegraphics[width=\columnwidth]{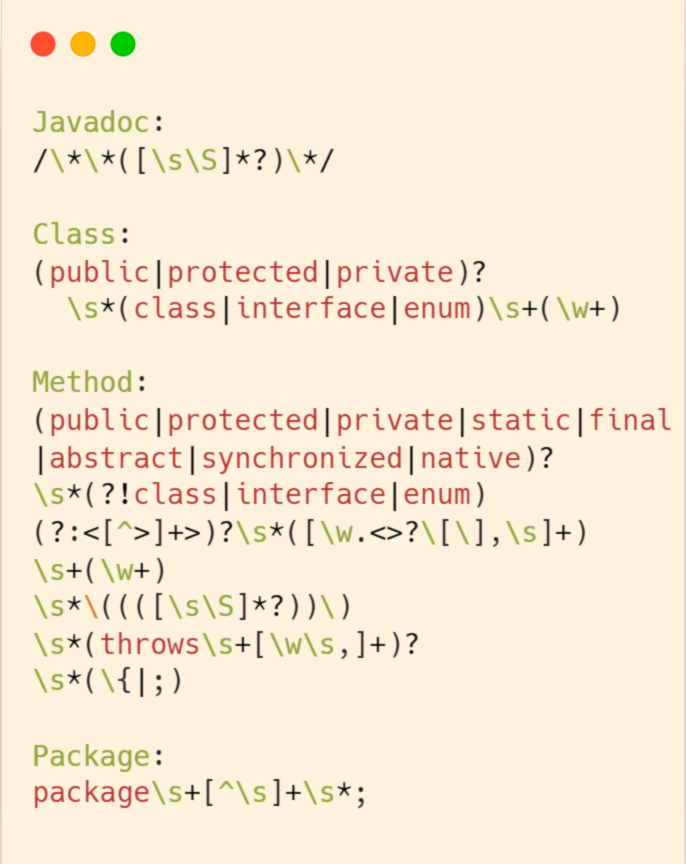}
    \caption{Regular Expressions for Extracting Javadoc Comments, Classes, Methods, and Packages from Java Source Code.}
    \label{fig:appendix_javadoc_regex}
\end{figure}

The regular expressions used to parse Java source files for Javadoc comments, package declarations, class/interface/enum declarations, and method/constructor signatures are detailed in Figure \ref{fig:appendix_javadoc_regex}.

\section{Evaluation Metrics}
\label{sec:appendix_metrics}
This section provides detailed definitions of the evaluation metrics used in our study.

\subsection{BLEU Score}
BLEU (Bilingual Evaluation Understudy) is an automated metric for evaluating machine translation by measuring n-gram overlap between a generated text $g$ and a reference text $r$. It incorporates a brevity penalty to discourage overly short outputs. Higher BLEU scores indicate a closer alignment with human-quality translations \cite{39}.

\subsection{ROUGE Score}
ROUGE (Recall-Oriented Understudy for Gisting Evaluation) is a set of metrics used for evaluating text summarization and machine translation. It compares a generated summary $g$ with one or more reference summaries $r$ by measuring overlap in units such as unigrams, bigrams, and the longest common subsequence (LCS) \cite{40}.

\paragraph{Notation}
\begin{itemize}
    \item $\text{Count}_{\text{match}}(u)$: Number of times a unigram $u$ from reference $r$ appears in the generated summary $g$.
    \item $\text{Count}(u', r)$: Total occurrences of unigram $u'$ in the reference summary $r$.
    \item $\text{Count}_{\text{match}}(b)$: Number of bigrams $b$ from reference $r$ that match in $g$.
    \item $\text{Count}(b', r)$: Total occurrences of bigram $b'$ in the reference summary $r$.
    \item $\text{LCS}(g, r)$: Length of the Longest Common Subsequence between generated summary $g$ and reference $r$.
    \item $L_r$: Total number of words in the reference summary $r$.
    \item $C_{\text{LCS}}$: Cumulative LCS over all sentence pairs between $g$ and $r$.
    \item $W_r$: Total word count across all sentences in the reference summary $r$.
\end{itemize}

\paragraph{ROUGE-1 (R1)}
Evaluates unigram overlap:
\begin{equation}
\text{R1} = \frac{\sum\limits_{u \in \text{Unigrams}(r)} \text{Count}_{\text{match}}(u)}
{\sum\limits_{u' \in \text{Unigrams}(r)} \text{Count}(u', r)}
\end{equation}

\paragraph{ROUGE-2 (R2)}
Evaluates bigram overlap:
\begin{equation}
\text{R2} = \frac{\sum\limits_{b \in \text{Bigrams}(r)} \text{Count}_{\text{match}}(b)}
{\sum\limits_{b' \in \text{Bigrams}(r)} \text{Count}(b', r)}
\end{equation}

\paragraph{ROUGE-L (RL)}
Calculates LCS normalized by reference length:
\begin{equation}
\text{RL} = \frac{\text{LCS}(g, r)}{L_r}
\end{equation}

\paragraph{ROUGE-Lsum (RLsum)}
Evaluates summary-level LCS similarity:
\begin{equation}
\text{RLsum} = \frac{C_{\text{LCS}}}{W_r}
\end{equation}

\section{Fine-Tuning Hyperparameters}
\label{sec:appendix_hyperparameters}

\subsection{LoRA Configuration}
We employed Low-Rank Adaptation (LoRA) for parameter-efficient fine-tuning. LoRA freezes the pre-trained model weights and injects trainable rank decomposition matrices into the Transformer architecture. The weight update is defined as:
\begin{equation}
W = W_0 + \Delta W = W_0 + BA
\end{equation}
where $W_0 \in \mathbb{R}^{d \times k}$ is the original weight matrix, and $B \in \mathbb{R}^{d \times r}$ and $A \in \mathbb{R}^{r \times k}$ are the trainable low-rank matrices, with rank $r \ll \min(d, k)$.

For our experiments, we set the rank to $r = 16$. The LoRA adaptation was applied to the following projection layers:
\begin{equation}
\begin{split}
\mathcal{T} = \{ q_{\text{proj}}, k_{\text{proj}}, v_{\text{proj}}, o_{\text{proj}}, \\
\quad \text{gate}_{\text{proj}}, \text{up}_{\text{proj}}, \text{down}_{\text{proj}} \}
\end{split}
\end{equation}
A scaling factor $\alpha = 16$ was used to moderate the magnitude of the weight updates.

\subsection{Training Configuration}
Table \ref{tab:appendix_training_config} provides an example of the training configuration used for the LLaMA-3.1-8B model. Similar hyperparameter settings were used for the other models, with minor adjustments where necessary.

\begin{table}[h!]
    \centering
    \caption{Example training configuration (LLaMA-3.1-8B)}
    \label{tab:appendix_training_config}
    \renewcommand{\arraystretch}{1.2}
    \footnotesize
    \begin{tabular}{l r}
        \toprule
        \textbf{Configuration} & \textbf{Value} \\
        \midrule
        Batch Size (Training)        & 8 \\
        Batch Size (Validation)      & 2 \\
        Gradient Accumulation Steps  & 4 \\
        Optimizer                    & AdamW \\
        Learning Rate                & \(2 \times 10^{-4}\) \\
        Evaluation Strategy          & steps \\
        Evaluation Steps             & 5 \\
        Linear Scheduler             & Yes \\
        Weight Decay                 & 0.01 \\
        Epochs                       & 5 \\
        \bottomrule
    \end{tabular}
\end{table}

\end{document}